\documentclass[aps,pre,epsf,superscriptaddress,amsmath,amssymb,amsfonts,twocolumn,showpacs]{revtex4-1}

\usepackage{graphicx}
\usepackage{epsfig}
\usepackage{dcolumn}
\usepackage{bm}
\usepackage{braket}
\usepackage{amsmath}
\usepackage{mathtools}
\usepackage{color}
\newcommand{\abs}[1]{\left| #1 \right|}

\begin{document}

\title{Quench Dynamics of Two One-Dimensional Harmonically\\ Trapped Bosons Bridging Attraction and Repulsion}

\author{L. Budewig}
\affiliation{Zentrum f\"{u}r Optische Quantentechnologien,
Universit\"{a}t Hamburg, Luruper Chaussee 149, 22761 Hamburg,
Germany}  
\author{S. I. Mistakidis}
\affiliation{Zentrum f\"{u}r Optische Quantentechnologien,
Universit\"{a}t Hamburg, Luruper Chaussee 149, 22761 Hamburg,
Germany}
\author{P. Schmelcher}
\affiliation{Zentrum f\"{u}r Optische Quantentechnologien,
Universit\"{a}t Hamburg, Luruper Chaussee 149, 22761 Hamburg,
Germany}\affiliation{The Hamburg Centre for Ultrafast Imaging,
Universit\"{a}t Hamburg, Luruper Chaussee 149, 22761 Hamburg,
Germany}

\date{\today}

\begin{abstract} 
We unravel the nonequilibrium quantum dynamics of two harmonically confined bosons in one spatial dimension when performing 
an interaction quench from finite repulsive to attractive interaction strengths and vice versa. 
A closed analytical form of the expansion coefficients of the time-evolved two-body wavefunction is derived, 
while its dynamics is determined in terms of an expansion over the postquench eigenstates. 
For both quench scenarios the temporal evolution is analyzed by inspecting the one- and two-body 
reduced density matrices and densities, the momentum distribution and the fidelity. 
Resorting to the fidelity spectrum and the eigenspectrum we identify the dominant eigenstates 
of the system that govern the dynamics. 
Monitoring the dynamics of the above-mentioned observables we provide signatures of the energetically higher-lying 
states triggered by the quench.    
\end{abstract}

%\pacs{Pacs numbers} 
\maketitle

\section{Introduction}

Ultracold quantum gases enable us to study a variety of many-body quantum phenomena due to an exquisite level of their control \cite {Lewenstein,Bloch}. 
The tremendous experimental progress, e.g. in terms of cooling and trapping techniques \cite{cooling}, grants the tunability of several system 
parameters. 
These include the size of the atomic sample \cite{Serwane,zurn,Wenz}, the shape 
and dimensionality of the external confinement \cite{Greiner_dim} as well as the interaction strength via Feshbach 
resonances \cite{Inouye,Chin}. 
For dilute ultracold quantum gases the interparticle interaction can be adequately approximated by a two-body contact interaction \cite{Huang,Derevianko}. 
Moreover, few-body systems due to their small number of degrees of freedom allow us to deepen our understanding of 
the system-specific microscopic mechanisms, the build up of quantum correlations and the few-to-many body crossover \cite{Blume}. 
A promising route to gain insight into such systems is to study their nonequilibrium quantum dynamics induced, for instance, 
via a quantum quench \cite{Polkovnikov,Langen}. 
In this context, the time-evolution is triggered after a sudden change of an intrinsic system parameter 
such as the interaction strength \cite{Kollath,Kollath1,Mistakidis,Mistakidis1}. 
The dynamics of such systems typically possesses a non-trivial dependence on all available degrees of freedom, 
rendering the analytical and even numerical treatment of few-body setups particularly challenging. 
Of course, the existence of analytical solutions is highly desirable since they provide complete information 
about the system under investigation. 

A prototype setup that can be analytically solved is the system of two atoms interacting in terms of an $s$-wave delta 
pseudopotential and being confined in a harmonic trap \cite{Busch,Farrell,Albeverio}. 
The energy spectrum of this system has also been verified experimentally \cite{Stoferle}. 
Furthermore, the stationary properties of two-bosons confined in isotropic \cite{Busch,Block,Cirone,Shea} and 
anisotropic \cite{Idziaszek1,Idziaszek2} harmonic 
potentials have been thoroughly investigated and a generalization of the pseudopotential when taking into account 
higher partial waves has been reported \cite{Stock,Idziaszek3}. 

However, the quantum dynamical evolution of the two harmonically trapped bosons when resorting to 
an analytical treatment is much less explored \cite{Ebert,Kehrberger,Ledesma,March}. 
For instance, the interaction effects of two atoms in a one-dimensional harmonic trap with time-dependent 
frequency have been studied \cite{Ebert}, showcasing that for a rapidly changing driving frequency 
the dynamics can be understood via a proper rescaling of the observables. 
The properties of the monopolar excitation have also been examined \cite{Ledesma} upon quenching the trapping 
frequency of the harmonic oscillator. 
In the same context and referring to very strong interactions the dynamical orthogonality of the postquench state 
with respect to the initial one has been found \cite{March}. 
Moreover, the dynamical generation of entanglement 
when considering low-energy collisions of the two atoms, each of them being initialized in a superposition of two counterpropagating wavepackets, 
has been investigated \cite{two_atom_ent}. 
It has been also recently shown that following an interaction quench from zero to infinitely strong 
interactions and vice versa a dynamical crossover from bosonic to fermionic properties is observed \cite{Kehrberger}.  
An important remark here is that in this latter investigation, the wavefunction of both the pre- and the postquench 
states is well-known due to the Bose-Fermi theorem \cite{Tonks,Girardeau}. 
However, the study of the interaction quench dynamics of the two-boson setup within the intermediate, either 
repulsive or attractive, interaction regime utilizing an exact analytical treatment \cite{Busch} remains to be addressed. 
Indeed for intermediate interaction strengths there is no closed form of the underlying basis states since 
the eigenenergies of the problem are given by the numerical solution of the corresponding 
transcendental equation. 
Such an investigation will permit us however to excite a variety of energetically higher-lying eigenstates and also unravel 
the role of the existing bound state by utilizing the avoided crossings occurring in the two-particle eigenspectrum. 
This knowledge might prove useful in future investigations for designing specific state 
preparation processes \cite{state_transfer,state_transfer1,state_transfer2,state_transfer3}. 

In this work we investigate the nonequilibrium dynamics of two harmonically trapped atoms 
in one-dimension by considering an interaction quench from repulsive to attractive interactions 
and vice versa. 
We first provide the analytical expression of the interacting two-body wavefunction for an arbitrary 
stationary eigenstate. 
Subsequently, the time-evolution of the two-body wavefunction in terms of the postquench 
eigenstates is determined. 
In particular, a closed analytical form of the corresponding expansion coefficients is derived. 
To explain the system's dynamical response upon an interaction quench we analyze the fidelity evolution 
and its spectrum \cite{Mistakidis,Jannis,Campbell}. 
The latter enables us to identify the predominantly participating eigenstates after the quench. 
The quench-induced spatial redistribution of the bosons, manifested as a breathing motion, is visualized 
by inspecting the time-evolution of the one- and two-body reduced density matrices \cite{Kehrberger}. 
We show that the spatial structures imprinted upon these density matrices in the course of the evolution 
signal the involvement of energetically higher-lying states \cite{few,few_attractive,momentum}.  
Moreover, the breathing motion of the cloud becomes evident in the time-evolution of the momentum distribution 
whose shape exhibits strong signatures of the higher-lying populated eigenstates. 
Finally, the dependence of the system's dynamical response either on the initial eigenstate for fixed postquench interaction strength 
or on the postquench interaction strength for the same initial state is thoroughly discussed. 
Remarkably enough, it is shown that the system's dynamical response exhibits a crossover from enhanced to weak response 
as a function of the postquench interaction strength. 
This alternating behavior is caused by the participation of the existing bound state to the postquench dynamics for strong attractions. 

This article is organized as follows. 
In Sec. \ref{theory}, we present our setup and discuss its stationary solutions and their time-evolution 
together with the relevant observables. 
In Sec. \ref{dynamics_positive}, we study the interaction quench dynamics from the repulsive to the attractive 
regime of interactions while in Sec. \ref{dynamics_negative} the reverse quench scenario 
(from attractive to repulsive interactions) is analyzed. 
We summarize our findings and discuss future perspectives in Sec. \ref{conclusions}. 
Appendix A delineates the convergence of our results in terms of the finite basis size used.

\section{Theoretical Framework}\label{theory} 

\subsection{Setup and Wavefunction Ansatz}  

We consider two identical harmonically trapped bosons located at ${x}_{1}$ and ${x}_{2}$ respectively in one spatial dimension. 
The interparticle interaction is modeled by a point-like $\delta$-potential of effective strength $\sqrt{2}g$ \cite{Kehrberger,Busch} 
which can be either repulsive ($g>0$) or attractive ($g<0$). 
The resulting Hamiltonian, rescaled in harmonic oscillator units i.e. $m=\hbar=\omega=1$, reads 
\begin{equation}
\label{eqn:Hamiltonian}
\begin{split}
H(x_1,x_2)=-{\frac{1}{2}}{\frac{\partial^2}{{\partial}{{x}_{1}}^2}}+{\frac{1}{2}}{x}_{1}^2-{\frac{1}{2}}
{\frac{\partial^2}{{\partial}{{x}_{2}}^2}}+{\frac{1}{2}}{x}_{2}^2\\+\sqrt{2}g{\delta}({x}_{1}-{x}_{2}),
\end{split}
\end{equation}
where the factor $\sqrt{2}$ has been introduced for later convenience. 
Introducing the center-of-mass (cm), $X=({x}_{1}+{x}_{2})/{\sqrt{2}}$, and relative (rel) coordinates, $x=({x}_{1}-{x}_{2})/{\sqrt{2}}$, 
reduces the two-particle problem to two effective single-particle problems. 
Consequently the Hamiltonian of Eq. (\ref{eqn:Hamiltonian}) is separated into its center-of-mass, $H_{cm}$, and relative coordinate parts, $H_{rel}$. 
Namely  
\begin{equation}
\label{eqn:Hamiltonian1}
H(X,x)=\underbrace{-{\frac{1}{2}}{\frac{\partial^2}{{\partial}{X}^2}}+{\frac{1}{2}}X^2}_{{{H}_{cm}}}
\underbrace{-{\frac{1}{2}}{\frac{\partial^2}{{\partial}{x}^2}}+{\frac{1}{2}}x^2+g{\delta}(x)}_{{{H}_{rel}}}.
\end{equation}
In this way, the total wavefunction of the system can be decomposed into its center-of-mass and relative eigenstates 
\begin{equation}
\label{eqn:Produktansatz}
{\Psi}({x}_{1},{x}_{2})={\Psi}_{cm}(X({x}_{1},{x}_{2})){\Psi}_{rel}(x({x}_{1},{x}_{2})).
\end{equation}
As it can be seen from Eq. (\ref{eqn:Hamiltonian}) the center-of-mass eigenstates are not affected by the interparticle interaction, $g$. 
Therefore, they correspond to the well-known harmonic oscillator eigenstates. 
In what follows we shall use the center-of-mass ground state ${\Psi}_{0;cm}(X)={\pi}^{-\frac{1}{4}}{e^{-\frac{X^2}{2}}}$. 
On the other hand, $H_{rel}$ is an effective single-particle problem referring to one particle in a 
harmonic trap with a delta-potential at the origin $x=0$. 
Since the bosonic exchange symmetry is reflected in the parity symmetry of the relative coordinate wavefunctions, we consider 
only the even eigenstates of $H_{rel}$. 
Thus, in the following, we denote the total wavefunction of Eq. (\ref{eqn:Produktansatz}) 
as ${\Psi}_{0;2\nu_i}({x}_{1},{x}_{2})$, where the quantum numbers $0$ and $2\nu_i$ (with $\nu_i=0,1,2,\dots$) stand for the center-of-mass ground state 
and the even eigenstates of $H_{rel}$ respectively. 
Note that from now on the index $i$ appearing in all quantities indicates that they refer to the initial (prequench) state 
of the system, while for the postquench (final) states 
we shall use the index $f$ [see also the discussion in Sec. \ref{quench protocol}].   

\begin{figure}[ht]
 	\centering
  	\includegraphics[width=0.46\textwidth]{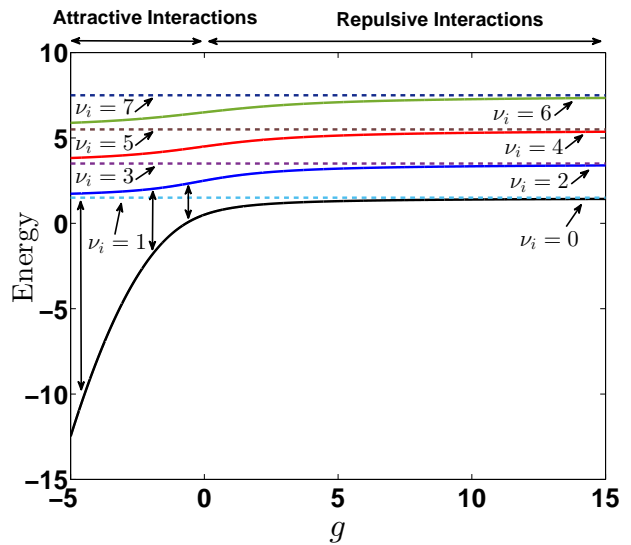}
     \caption{Energy spectrum as a function of the interaction strength, $g$, of two-bosons confined in an one-dimensional harmonic trap. 
Both the eigenenergies of the symmetric ($\nu_i=0,2,4,\dots$) interaction-dependent and antisymmetric ($\nu_i=1,3,\dots$) interaction-independent 
eigenstates of the relative motion for two-bosons are depicted. 
The double arrows indicate the energy gap between the $\nu_i=0$ and $\nu_i=2$ energy levels at different interaction strengths. } 
 	\label{fig:E}
\end{figure}

We employ as an ansatz for the even eigenstates of $H_{rel}$ a time-independent 
superposition in terms of the non-interacting even eigenstates of the harmonic 
oscillator. 
The latter are ${\varphi}_{2n}(x)=(1/\pi^{1/4}\sqrt{2^{2n}(2n)!})H_{2n}(x)e^{-x^2/2}$, with $H_{2n}(x)$ denoting 
the corresponding Hermite polynomials. 
Then, the expansion of the relative wavefunction \cite{Busch} reads 
\begin{equation}
\label{eqn:w}
{\Psi}_{2\nu_i;rel}(x)=A^{2\nu_i}\sum_{0 \le n \le \infty}\frac{{\varphi}_{2n}^*(0)}{{E}_{2n}-{E}_{2\nu_i}}{\varphi}_{2n}(x),
\end{equation}
where the index $n$ refers to the energetic order of the non-interacting harmonic oscillator eigenstates. 
Moreover, $E_{2n}$ and $E_{2\nu_i}$ are the energies of the even eigenstates of the non-interacting and 
interacting (relative coordinates) case respectively. 
The normalization constant $A^{2\nu_i}$ can be expressed \cite{Cirone} in the following closed form 
\begin{equation}
 A^{2\nu_i}=\sqrt{\frac{4\Gamma (\frac{1}{2}-\epsilon_i)}{\Gamma(-\epsilon_i)[\psi(\frac{1}{2}-\epsilon_i)-\psi(-\epsilon_i)]}}.\label{coef_closed_form}
\end{equation}
In this expression $\epsilon_i=\frac{E_{2\nu_i}}{2}-\frac{1}{4}$, while $\Gamma$ and $\psi$ refer to the gamma and 
digamma functions \cite{Gradshteyn,Abramowitz,Andrews} respectively.  
We remark that the eigenstates [see Eq. (\ref{eqn:w})] can otherwise be expressed in terms of the confluent 
hypergeometric function $U$ and the gamma function $\Gamma$ \textbf{\cite{Ebert,two_atom_ent}} as 
\begin{equation}
\label{eqn:S!}
\begin{split}
{\Psi}_{2\nu_i;rel}(x)=&\frac{{A}^{2\nu_i}}{2\sqrt{\pi}}{\Gamma}(-\epsilon_i) U(-\epsilon_i,\frac{1}{2},x^{2})e^{-\frac{x^2}{2}}.
\end{split}
\end{equation} 
Furthermore, the energy-eigenvalues of the even eigenstates of $H_{rel}$ can be determined for a fixed value of the interaction 
strength by numerically solving the 
following transcendental equation \cite{Busch}   
\begin{equation}
\label{eqn:e}
\frac{\Gamma(-\frac{{E}_{2\nu_i}}{2}+\frac{3}{4})}{\Gamma(-\frac{{E}_{2\nu_i}}{2}+\frac{1}{4})}=-\frac{g}{2}. 
\end{equation} 
Figure \ref{fig:E} presents the first few lowest-lying eigenenergies of $H_{rel}$ obtained via Eq. (\ref{eqn:e}). 
As expected, in the non-interacting limit, $g=0$, the eigenspectrum of $H_{rel}$ corresponds to the 
single-particle eigenspectrum of the harmonic oscillator possessing equidistant eigenenergies 
$\nu_i+1/2$ with $\nu_i=0,1,2,\dots$ indexing the energy levels. 
However, for a finite value of $g$, being either positive or negative, the energy spectrum is altered. 
In particular, in the case of $g\neq0$ the odd levels $E_{2\nu_i+1}$ are not affected by the interaction potential $\delta(x)$ since 
their eigenstates always exhibit a node at $x=0$. 
On the contrary, the even states acquire an increasing (decreasing) energy $E_{2\nu_i}$ for $g>0$ ($g<0$).  
Ultimately, each even level approaches energetically the next (previous) upper (lower) odd level at $g\to\infty$ ($g\to-\infty$) thus 
forming a doublet spectrum which is characteristic for double-well potentials \cite{double_well,double_well1,few_dw}.  

\subsection{Basic Observables in a Stationary State}

Let us next briefly introduce a few basic observables when the system resides in a stationary eigenstate characterized by the quantum number $\nu_i$. 
These observables will be subsequently employed for the interpretation of the interaction quench dynamics. 
The one-body reduced density matrix \cite{Naraschewski,Sakmann_cor}, $\rho^{(1)}(x_1,x_1')$, provides the probability to find one boson, due to its wave nature, 
simultaneously at positions $x_1$ and $x_1'$ 
\begin{equation}
\label{eqn:RHO}
\begin{split}
{\rho}^{(1)}({x}_{1},{x}_{1}')&={\int_{-\infty}^{\infty}{d{x}_{2}{\Psi}_{0;2\nu_i}({x}_{1},{x}_{2}){\Psi}_{0;2\nu_i}^{*}({x}_{1}',{x}_{2})}}\\
&=\sum_{0 \le k \le \infty}{\lambda}_{k}^{2\nu_i}{\beta}_{k}^{2\nu_i}({x}_{1}){\beta}_{k}^{2\nu_i*}({x}_{1}').
\end{split}
\end{equation} 
We remark that its diagonal i.e. $x'_1=x_1$ is the single-particle density $\rho^{(1)}(x_1)\equiv \rho^{(1)}(x_1,x_1'=x_1)$. 
In the second line of Eq. (\ref{eqn:RHO}) we have also introduced the decomposition of ${\rho}^{(1)}({x}_{1},{x}_{1}')$ in terms of its corresponding 
eigenfunctions, the so-called natural orbitals $\beta_k^{2\nu_i}$, and eigenvalues ${\lambda}_{k}^{2\nu_i}$ termed natural populations \cite{two_atom_ent}. 
The index $k$ labels the $k$-th natural orbital and natural population respectively of the initial eigenstate with quantum number $2\nu_i$. 
The natural populations are ${\lambda}_{k}^{2\nu_i}=\mathcal{N}^{-2}(k)\frac{2^{k}k!}{\pi^{\frac{1}{2}}}$ with 
$\mathcal{N}^{-2}(k)=\frac{2^{-k}}{k!}\pi^{1/2}\sum_{0<n<\infty}2^{-2n} {{2n}\choose{k}}\abs{\frac{A^{2\nu_i}\phi_{2n}^*(0)}{E_{2n}-E_{2\nu_i}}}^2$.
Regarding the stationary eigenstates of quantum number $2\nu_i$ [see Eq. (\ref{eqn:w}) or (\ref{eqn:S!})] it can be easily shown that  
\begin{equation}
\label{eqn:g}
\begin{split}
%{n}^{2\nu_i}_{k}({x_1})=\mathcal{N}(k)\sum_{0 \le n \le \infty} 0 \le n \le \infty    \frac{2^{-2n}}{\sqrt{(2n)!}}\binom{2n}{k}\\ \times \frac{A^{2\nu_i}{\varphi}_{2n}^*(0)}{{E}_{2n}-{E}_{2\nu_i}}{H}_{2n-k}(x_1)e^{-\frac{x_1^2}{2}},
{\beta}^{2\nu_i}_{k}({x_1})=\mathcal{N}(k)\sum_{\substack{0 \le n \le \infty \\ {\rm with}~ k\le2n}}\frac{2^{-2n}}{\sqrt{(2n)!}}\binom{2n}{k}\\ \times \frac{A^{2\nu_i}{\varphi}_{2n}^*(0)}{{E}_{2n}-{E}_{2\nu_i}}{H}_{2n-k}(x_1)e^{-\frac{x_1^2}{2}},
%{\substack{j\in\{1,\dots,m^{\sigma}\} \\ {\rm with}~ l_j=1}}
\end{split}
\end{equation}
with $n$ indexing the energetic order of the non-interacting harmonic oscillator eigenstates. 
The normalization constant $\mathcal{N}(k)$ ensures that the natural orbitals are normalized to unity. 
The corresponding momentum distribution \cite{Kehrberger,two_atom_ent,Mistakidis_driven} of the single-particle density matrix reads 
\begin{equation}
n({p}_{1})=\sum_{0 \le k \le \infty}{{\lambda}}_{k}^{2\nu_i}|{{\beta}}_{k}^{2\nu_i}({p}_{1})|^2,
\end{equation} 
where $k$ labels the natural orbitals. 
In this expression we have used that 
\begin{equation}
\begin{split}
{{\beta}}_{k}^{2\nu_i}({p}_{1})=(-1)^{-\frac{3k}{2}}\mathcal{N}(k)\sum_{\substack{0 \le n \le \infty \\ {\rm with}~ k\le2n}}(-1)^{3n}\frac{2^{-2n}}{\sqrt{(2n)!}}\binom{2n}{k}\\ 
\times \frac{A^{2\nu_i}{\varphi}_{2n}^*(0)}{{E}_{2n}-{E}_{2\nu_i}}{H}_{2n-k}({p}_{1})e^{-\frac{{p}_{1}^2}{2}}.
\end{split}
\end{equation} 
Since we are also interested in examining the occurrence of two-body correlations into the system 
we will, furthermore, explore the diagonal of the two-body reduced density matrix \cite{Sakmann_cor}  
\begin{equation}
{\rho}^{(2)}({x}_{1},{x}_{2})=|{\Psi}_{0;2\nu_i}({x}_{1},{x}_{2})|^2. 
\end{equation}
This quantity refers to the probability of finding two atoms located at positions 
$x_1$ and $x_2$ respectively.

\subsection{Quench Protocol and Time-Evolution of the Wavefunction}\label{quench protocol}

We prepare our system in a stationary eigenstate ${\Psi}_{0;2\nu_i}({x}_{1},{x}_{2})$ at 
interaction strength $g_{i}$. 
To trigger the dynamics we then quench the interaction strength instantaneously from the value $g_{i}$ 
to $g_{f}$. 
We remark that the indices $i$ and $f$ in every quantity denote that it refers to the initial (prequench) and 
final (postquench) state of the system respectively. 
The postquench system is characterized by its stationary eigenstates ${\Psi}_{0;2\nu_f}({x}_{1},{x}_{2})$ with 
energy eigenvalues ${E}_{2\nu_f}+\frac{1}{2}$. 
Note that the value $\frac{1}{2}$ stems from the corresponding center-of-mass ground state. 
Subsequently, the time-evolving state at time $t$ is described by the expansion over 
the eigenstates of the postquench system with their time-dependent phase 
\begin{equation}
\label{eqn:L}
\begin{split}
&\Psi_{0;2\nu_f}(x_1,x_2;t)\\&=\sum_{0 \le \nu_f \le \infty}e^{-it({E}_{2\nu_f}+\frac{1}{2})}{C}_{2\nu_f;2\nu_i}{\Psi}_{0;2\nu_f}({x}_{1},{x}_{2}).
\end{split}
\end{equation} 
Here, ${C}_{2\nu_f;2\nu_i}=\braket{\Psi_{0;2\nu_f}(x_1,x_2)|\Psi_{0;2\nu_i}(x_1,x_2)}=\braket{{\Psi}_{2\nu_f;rel}(x)|{\Psi}_{2\nu_i;rel}(x)}$ denote 
the overlap coefficients between the initial eigenstate of the system and the final one.  
Using the relative eigenstates given by Eq. (\ref{eqn:w}) and the orthogonality of the harmonic oscillator functions ${\varphi}_{2n}(x)$ we can 
show that
\begin{equation}
\begin{split}
{C}_{2\nu_f;2\nu_i}=\frac{A^{2\nu_i}A^{2\nu_f}}{{E}_{2\nu_i}-{E}_{2\nu_f}}&[\sum_{0 \le n \le \infty}\frac{{\varphi}_{2n}(0){\varphi}_{2n}^*(0)}{{E}_{2n}-{E}_{2\nu_i}}\\
&-\sum_{0 \le n \le \infty}\frac{{\varphi}_{2n}(0){\varphi}_{2n}^*(0)}{{E}_{2n}-{E}_{2\nu_f}}].
\end{split}
\end{equation} 
Most importantly, utilizing the relation $\sum_{0 \le n \le \infty}\frac{{\varphi}_{2n}(0){\varphi}_{2n}^*(0)}{{E}_{2n}-{E}_{2\nu_i}}=
\frac{\Gamma(-\frac{{E}_{2\nu_i}}{2}+\frac{1}{4})}{2\Gamma(-\frac{{E}_{2\nu_i}}{2}+\frac{3}{4})}$ [see also Eq. (\ref{eqn:e})], the overlap coefficients 
acquire the closed form 
\begin{equation}
\label{eqn:C}
{C}_{2\nu_f;2\nu_i}=\frac{A^{2\nu_i}A^{2\nu_f}}{g_{i}g_{f}}\frac{{g}_{i}-{g}_{f}}{{E}_{2\nu_i}-{E}_{2\nu_f}}. 
\end{equation} 
In this expression the normalization constants $A^{2\nu_i}$, $A^{2\nu_f}$ can be calculated analytically [see also Eq. (\ref{coef_closed_form})], whereas the 
energies of the even energy levels of the pre- and postquench Hamiltonian $E_{2\nu_i}$ and $E_{2\nu_f}$ are determined 
via solving the transcendental Eq. (\ref{eqn:e}) \cite{Busch,Cirone}. 
In this way, we have obtained a closed analytical form for the expansion coefficients of the time-dependent two-body 
wavefunction $\Psi_{0;2\nu_f}(x_1,x_2;t)$. 
Therefore in order to calculate the time-evolution of $\Psi_{0;2\nu_f}(x_1,x_2;t)$ one needs to 
numerically determine Eq. (\ref{eqn:L}) being an infinite summation of postquench eigenstates characterized by the quantum numbers $\nu_f$. 
Of course, in practice this infinite summation is truncated to a finite one given that the values of all system's 
observables have been converged with respect to a further adding of eigenstates. 
This truncation procedure is showcased in Appendix A for some observables. 
We finally note that it can be shown that $\lim_{\substack{g_i\to g_f\\ 2\nu_i \to 2\nu_f}}C_{{2\nu}_f;{2\nu}_i}=1$.

\subsection{Time-Evolution of Observables}

Having at hand the postquench system's wavefunction [Eq. (\ref{eqn:L})] we can calculate 
the time-evolution of the observables of interest. 
Accordingly, the time-evolution of the reduced single-particle density matrix reads  
\begin{equation}
{\rho}^{(1)}({x}_{1},{x}_{1}';t)=\sum_{0 \le k \le \infty}{\lambda}_{k} {\beta}_{k}({x}_{1};t){\beta}^{*}_{k}({x}_{1}';t).
\end{equation} 
Here, ${\lambda}_{k}=\bigg(\sum_{0\le\nu_f\le\infty}\sqrt{\lambda_k^{2\nu_f}}\bigg)^2$ denotes the $k$-th eigenvalue of ${\rho}^{(1)}({x}_{1},{x}_{1}';t)$ and 
$\lambda_k^{2\nu_f}$ refers to the $k$-th natural population of the $2\nu_f$ postquench eigenstate at interaction strength $g_f$. 
Moreover, ${\beta}_{k}({x}_{1};t)$ is the $k$-th natural orbital being the $k$-th eigenfunction of ${\rho}^{(1)}({x}_{1},{x}_{1}';t)$. 
The ${\beta}_{k}({x}_{1};t)$ can be expressed in terms of the natural orbitals ${\beta}_{k}^{2\nu_f}(x_1)$ of the corresponding stationary eigenstates (characterized by 
the quantum number $2\nu_f$) at interaction strength $g_f$ as  
\begin{equation}
{\beta}_{k}(x_1;t)=\sum_{0 \le \nu_f \le \infty}e^{-\imath{E}_{2\nu_f}t}{C}_{2\nu_f;2\nu_i}\beta^{2\nu_f}_{k}(x_1).
\end{equation} 
The momentum distribution during the dynamics \cite{Jannis,Mistakidis_driven,two_atom_ent} reads 
\begin{equation}
n({p}_{1};t)=\sum_{0 \le k \le \infty}{{\lambda}}_{k}|{{\beta}}_{k}({p}_{1};t)|^2, \label{momentum_evolution}
\end{equation} 
with ${\beta}_{k}({p}_{1};t)=\sum_{0 \le \nu_f \le \infty}e^{-\imath{E}_{2\nu_f}t}{C}_{2\nu_f;2\nu_i}{\beta}_{k}^{2\nu_f}({p}_{1})$. 
Moreover, the time-evolution of the two-body reduced density matrix is   
\begin{equation}
{\rho}^{(2)}({x}_{1},{x}_{2};t)=|{\Psi}_{0;2\nu_f}({x}_{1},{x}_{2};t)|^2.\label{two_body_evol}
\end{equation}
As we have already argued above ${\rho}^{(2)}({x}_{1},{x}_{2};t)$ provides the probability to detect two atoms 
at a fixed time instant $t$ at positions $x_1$ and $x_2$, respectively. 
Finally, the overlap between the initial (stationary state) and the time-evolving 
wavefunction yields the fidelity \cite{Gorin,Venuti,Campbell} of the system 
\begin{equation}
\begin{split}
{F}(t)&=\abs{\braket{{\Psi}_{2\nu_i;rel}(x;t=0)|{\Psi}_{2\nu_f;rel}(x;t)}}\\
&=|\sum_{0 \le \nu_f \le \infty} e^{-i{E}_{2\nu_f}t}|{C}_{2\nu_f;2\nu_i}|^2|. \label{eq:fidelity}
\end{split}
\end{equation}
This quantity provides a time-resolved measure for the effect of the quench onto the system and therefore 
dictates its dynamical response following the quench \cite{Mistakidis,Mistakidis1,Mistakidis7,Mistakidis6,Jannis,Campbell}. 
We remark that for the fidelity calculation only the relative coordinate states contribute since the center-of-mass is unperturbed. 
${F}(t)$ takes values from zero (the two states are orthogonal) to unity (the two states are the same). 
Since the fidelity offers a measure for the system's dynamical response its spectrum 
\begin{equation}
\begin{split}
&{F}(\omega)=\frac{1}{\sqrt{2\pi}}\int_{-\infty}^{\infty}{dt|{F}(t)|^2e^{\imath{\omega}t}}\\\
&=\sqrt{2\pi}\sum_{0 \le \nu_f,\nu_h \le \infty}|{C}_{2\nu_f;2\nu_i}|^2|{C}_{2\nu_h;2\nu_i}|^2\delta(\omega-{\omega}_{2\nu_f;2\nu_h}), \label{eq:fidelity_spectrum} 
\end{split}
\end{equation}
provides information about the frequencies of the quench-induced modes. 
Here, ${\omega}_{2\nu_f;2\nu_h}\equiv{E}_{2\nu_f}-{E}_{2\nu_h}$ refers to the energy difference between two 
even energy levels (denoted by $2\nu_f \equiv \nu'_f$ and $2\nu_h \equiv \nu'_h$) of the postquench system. 
The contribution of each frequency $\omega_{2\nu_f;2\nu_h}$ to ${F}(t)$ can be quantified via the corresponding overlap 
coefficients, $\abs{{C}_{2\nu_f;2\nu_i}}^2$ and $\abs{{C}_{2\nu_h;2\nu_i}}^2$, of the two involved postquench states 
i.e. ${\Psi}_{2\nu_f;rel}(x)$ and ${\Psi}_{2\nu_h;rel}(x)$ with the prequenched (initial) state ${\Psi}_{2\nu_i;rel}(x)$. 
Note finally that since we consider the total wavefunction of the system to be normalized to unity, 
then $\sum_{\nu_f=0}^{\infty}\abs{{C}_{2\nu_f;2\nu_i}}^2=1$ holds. 

Having introduced the basic formalism and observables for the time-evolution of the two-boson system we next proceed to 
the description of the corresponding interaction quench-induced dynamics. 
In particular, we shall mainly analyze two different quench scenarios: 
an interaction quench from the repulsive towards the attractive side of interactions 
[see Section \ref{dynamics_positive}] and vice versa [see Section \ref{dynamics_negative}].

\section{Quench Dynamics from Repulsive to Attractive Interactions}\label{dynamics_positive}

\begin{figure}[ht]
 	\centering
  	\includegraphics[width=0.48\textwidth]{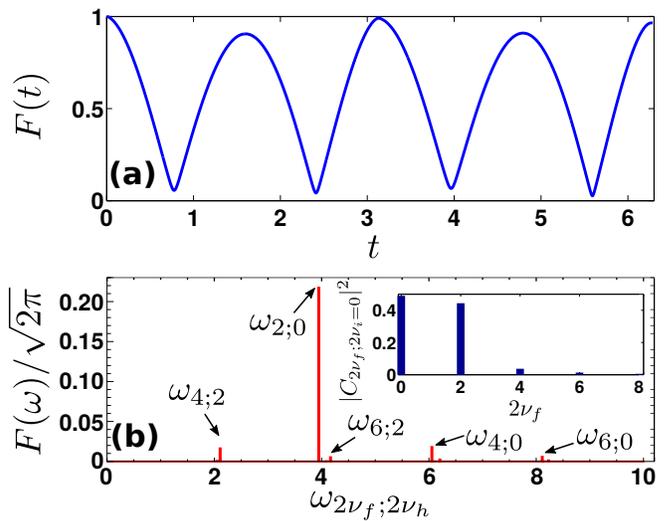}
     \caption{(a) Time-evolution of the fidelity for an interaction quench from $g_i=2$ to $g_{f}=-2$. 
     (b) The corresponding fidelity spectrum $F(\omega)$. 
     The inset shows the square of the overlap coefficients, $\abs{{C}_{2\nu_f;2\nu_i=0}}^2$, between the initial (ground, $2\nu_i=0$) state 
     of the system at $g_i=2$ and each of the first five ($\nu'_f\equiv2\nu_f=0,2,\dots,8$) even eigenstates of the postquench, $g_f=-2$, system. 
In all cases the system consists of two-bosons confined in an one-dimensional harmonic oscillator and it is initialized in the ground state of $g_i=2$.} 
 	\label{fig:F5}
\end{figure} 

We first examine the characteristics of the nonequilibrium dynamics of two harmonically trapped bosons when 
considering a sudden change (quench) of their interaction strength from repulsive to attractive interactions. 
In particular, the system is initially prepared in its ground state $E_{2\nu_i=0}$ for $g_i=2$ and at $t=0$ an interaction 
quench towards $g_f=-2$ is performed enforcing the system to evolve.    

\begin{table}
  
 \begin{center}
  \begin{tabular}{|c| c| c | c | c | c | c |c |c |} 
  \hline
  $g_i$ & $g_f$ & $\Delta E_{2;0}$ & $\Delta E_{4;0}$ & $\Delta E_{6;0}$ & $\Delta E_{4;2}$ & $\Delta E_{6;2}$ & $\Delta E_{8;2}$ & $\Delta E_{8;0}$\\ [0.5ex] 
  \hline\hline
   2 & -2 & 3.95 & 6.06&  8.11& 2.11& 4.17 & 6.20 & 10.15  \\ 
  \hline
   -2 & 2 & 1.85 & 3.78& 5.73& 1.93& 3.88 & 5.85 & 7.70\\ [1ex]
  \hline
 
 \end{tabular}
 \caption {Energy differences $\Delta E_{2\nu_f;2\nu_h}=E_{2\nu_f}-E_{2\nu_h}$ between two distinct even eigenstates, $2\nu_f$ and $2\nu_h$ respectively, of the postquench system.  
 The system of two-bosons is prepared in its ground state of interaction strength $g_i$ and at $t=0$ the interaction strength is quenched 
 to $g_f$.}\label{table}
 \end{center}
 
 \end{table}

\begin{figure}[ht]
 	\centering
  	\includegraphics[width=0.48\textwidth]{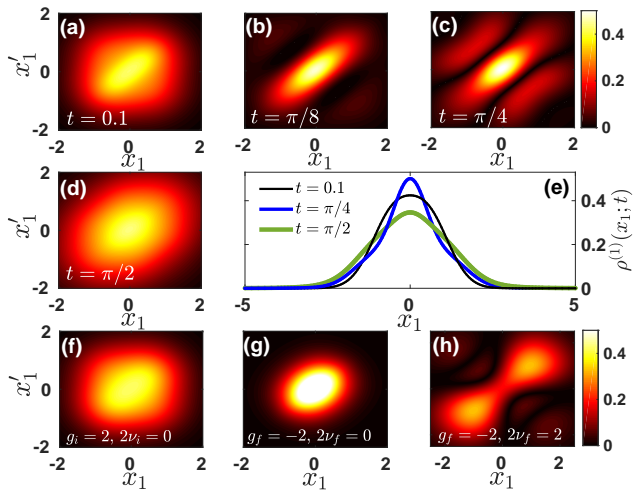}
     \caption{(a)-(d) Time-evolution of the one-body reduced density matrix, $\rho^{(1)}(x_1,x_1';t)$, at different time-instants (see legends) following an interaction quench 
	from $g_i=2$ to $g_f=-2$. 
	(e) Snapshots of the one-body density, $\rho^{(1)}(x_1,x_1'=x_1;t)$, at distinct times (see legend) during the evolution. 
	$\rho^{(1)}(x_1,x_1';t)$ for the ground state of (f) $g_i=2$ and (g) $g_f=-2$ and (h) the second excited state ($\nu'_f\equiv 2\nu_f=2$) of $g_f=-2$. 
	In all cases we consider two harmonically trapped bosons in one spatial dimension prepared in their corresponding ground state with $g_i=2$.} 
 	\label{fig:SDM5}
\end{figure} 

To inspect the system's overall dynamical response we utilize the fidelity evolution $F(t)$ [see also Eq. (\ref{eq:fidelity})], 
which essentially provides the overlap between the time evolved and the initial state of the system \cite{Mistakidis6,Mistakidis7,Gorin}. 
As it can be seen in Fig. \ref{fig:F5} (a), $F(t)$ deviates significantly from unity indicating a noteworthy perturbation 
of the system from its initial state. 
In particular, $F(t)$ exhibits an oscillatory behavior involving several frequencies. 
During this oscillatory motion of $F(t)$ the system is driven away from its initial state when $F(t)\ll1$ and returns 
close to it within the time-intervals where $F(t)\approx1$. 
The involved frequencies can be quantified via the corresponding fidelity spectrum $F(\omega)$ [see also Eq. (\ref{eq:fidelity_spectrum})] which 
is shown in Fig. \ref{fig:F5} (b). 
Indeed, we observe the appearance of five frequency peaks located at $\omega_{4;2}\approx2.11$, $\omega_{2;0}\approx3.95$, $\omega_{6;2}\approx4.17$, 
$\omega_{4;0}\approx6.06$ and $\omega_{6;0}\approx8.11$ respectively. 
Recall that we use the notation $\omega_{2\nu_f;2\nu_h}$, where $2\nu_f$ and $2\nu_h$ denote two distinct even energy levels of the postquench system. 
Moreover in order to assign each frequency peak appearing in $F(\omega)$ to a corresponding energy difference of two 
eigenstates of the postquench system we perform the following analysis. 
We first determine the overlap coefficients $\abs{{C}_{2\nu_f;2\nu_i=0}}^2$ between the prequenched (initial) state, here $2\nu_i=0$ at $g_i=2$, and different 
even states ($2\nu_f=0,2,4,\dots$ at $g_f=-2$) of the postquench system. 
The values of the corresponding coefficients $\abs{{C}_{2\nu_f;2\nu_i=0}}^2$ are presented in the inset of Fig. \ref{fig:F5} (b). 
As shown only the first five even eigenstates, i.e. $2\nu_f=0,2,\dots,8$, have a non-negligible overlap with the initial state and especially 
the $2\nu_f=0$ and $2\nu_f=2$ possess the dominant overlap. 
Combining the above knowledge in terms of the $\abs{{C}_{2\nu_f;2\nu_i=0}}^2$ with the corresponding energy difference between two distinct 
eigenstates at $g_f$, see Fig. \ref{fig:E} and also Table \ref{table}, we conclude upon the assignment of each frequency peak 
in $F(\omega)$ to a transition among two specific even final eigenstates e.g. $2\nu_f$ and $2\nu_h$. 
In particular, only transitions between the eigenstates $2\nu_f$ and $2\nu_h$ which have a non-negligible overlap with the 
initial state (i.e. $\abs{{C}_{2\nu_f;2\nu_i=0}}^2\neq0$ and $\abs{{C}_{2\nu_h;2\nu_i=0}}^2\neq0$) are permitted. 
Then, the value of $\omega_{2\nu_f;2\nu_h}$ occurring in $F(\omega)$ should be the same with $\Delta E_{2\nu_f;2\nu_h}=E_{2\nu_f}-E_{2\nu_h}$ 
determined from the two-body eigenspectrum [Fig. \ref{fig:E}]. 
Evidently, $\omega_{2;0}$ is the dominant participating frequency in $F(t)$ since it possesses the maximum amplitude in $F(\omega)$.  

\begin{figure}[ht]
 	\centering
  	\includegraphics[width=0.46\textwidth]{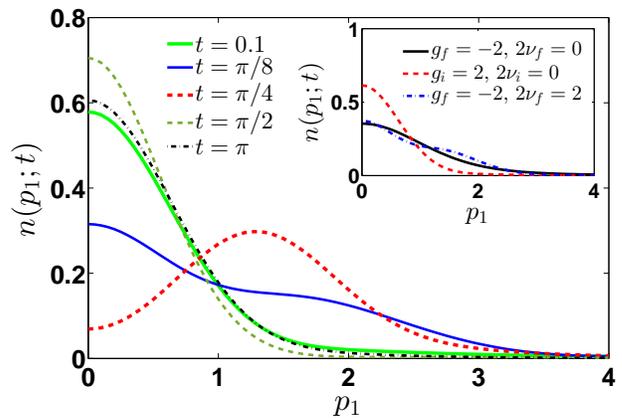}
     \caption{Momentum distribution $n(p_1;t)$ of the one-body density matrix at specific time-instants (see legend) of the evolution following an interaction quench from 
${g}_{i}=2$ to ${g}_{f}=-2$. For simplicity, only positive momenta ${p}_{1}$ are depicted. 
The inset shows $n(p_1,0)$ for certain even eigenstates ($\nu'_f\equiv 2\nu_f$) referring to interactions before or after the quench (see legend). 
The system consists of two-bosons initialized in the ground state of an one-dimensional harmonic trap with $g_i=2$.} 
 	\label{fig:P5}
\end{figure}

\subsection{Dynamics on the Single-Particle Level} 

To visualize the spatially resolved dynamics of the two-bosons on the single-particle level we next resort to the evolution of the one-body 
reduced density matrix, $\rho^{(1)}(x_1,x_1';t)$, following the interaction quench, see Fig. \ref{fig:SDM5}. 
We remark that this quantity shows the spatial distribution of one boson while its diagonal, i.e. $x_1=x_1'$, corresponds to the one-body density 
of the system. 
Starting from the ground state of $g_i=2$ [Fig. \ref{fig:SDM5} (f)] we observe that $\rho^{(1)}(x_1,x_1';t)$ deforms at the initial stages of 
the dynamics [see Fig. \ref{fig:SDM5} (a)] developing a two-hump structure along its anti-diagonal ($x_1=-x_1'$) for later times [see Fig. \ref{fig:SDM5} (b) and (c)] and 
returns almost back to its initial shape for even longer times [see Fig. \ref{fig:SDM5} (d)]. 
Then, $\rho^{(1)}(x_1,x_1';t)$ performs a similar to the aforementioned dynamics in the course of the evolution in an almost periodic manner (not shown here for brevity reasons). 
The above-described almost periodic deviation in time of the system from its initial state is essentially reflected by the oscillatory behavior of its $F(t)$ 
as discussed in Fig. \ref{fig:F5} (a). 
Most importantly, the two-hump structure appearing in the anti-diagonal of $\rho^{(1)}(x_1,x_1';t)$ [Figs. \ref{fig:SDM5} (b), (c)] during the evolution is caused by the 
superposition of the ground and the energetically higher-lying excited states of the postquench system \cite{few,momentum,few_dw}, see also the inset of Fig. \ref{fig:F5} (b). 
The occurrence of such a superposition can also be inferred by inspecting the $\rho^{(1)}(x_1,x_1';0)$ of these eigenstates, see 
e.g. Figs. \ref{fig:SDM5} (g) and (h) where $\rho^{(1)}(x_1,x_1';0)$ of the ground ($2\nu_f=0$) and the second excited ($2\nu_f=2$) states are depicted. 
Turning to the diagonal of $\rho^{(1)}(x_1,x_1'=x_1;t)\equiv\rho^{(1)}(x_1;t)$, i.e. the one-body density of the system [see Fig. \ref{fig:SDM5} (e)], we 
can deduce that the cloud undergoes an expansion and contraction dynamics which is a manifestation, of course, of its underlying 
breathing motion \cite{Abraham_breathing,Abraham_breathing1}. 

As a next step and in order to analyze the motion of the two-atoms in momentum space \cite{two_atom_ent,Kehrberger} we employ the time-evolution of the momentum distribution 
of the one-body density matrix $n(p_1;t)$ [see Eq. (\ref{momentum_evolution})]. 
This quantity is directly experimentally accessible via time-of-flight measurements \cite{Bloch}. 
Figure \ref{fig:P5} presents $n(p_1;t)$ for distinct time-instants of the interaction quench dynamics from the repulsive towards the attractive 
regime of interactions. 
Note that for simplicity only positive momenta, i.e. $p_1>0$, are depicted since $n(p_1;t)$ is symmetric with respect to $p_1=0$. 
As expected $n(p_1;t)$ shows a behavior which is reminiscent of the corresponding of the one-body density $\rho^{(1)}(x_1;t)$, 
compare Fig. \ref{fig:SDM5} (f) and Fig. \ref{fig:P5}. 
In this way, the overall expansion and contraction of $n(p_1;t)$ during dynamics essentially visualizes the breathing motion in momentum space \cite{Jannis,momentum}. 
Initially, e.g. at $t=0.1$, $n(p_1;t)$ exhibits a peak around $p_1=0$ resembling also $n(p_1;t=0)$, see the inset of Fig. \ref{fig:P5}. 
As time evolves (e.g. $t=\pi/8$) a strong reduction of the zero-momentum peak 
is observed and a new maximum at finite momenta occurs, e.g. at $t=\pi/4$ in the vicinity of $p_1\approx1.2$. 
The appearance of this peak at finite $p_1$ is mainly caused due to the superposition of the two-boson state in terms of the ground ($2\nu_f=0$) and the 
second ($2\nu_f=2$) excited states for $g_f=-2$ [see also the inset of Fig. \ref{fig:P5}]. 
For longer times, e.g. at $t=\pi/2$, $n(p_1;t)$ turns back to its original shape having a maximum in the neighborhood of $p_1=0$. 
As we argued above the motion is almost periodic, and therefore $n(p_1;t)$ undergoes a similar to the above-mentioned dynamics in the 
course of the evolution (not shown here).   

\begin{figure}[ht]
 	\centering
  	\includegraphics[width=0.46\textwidth]{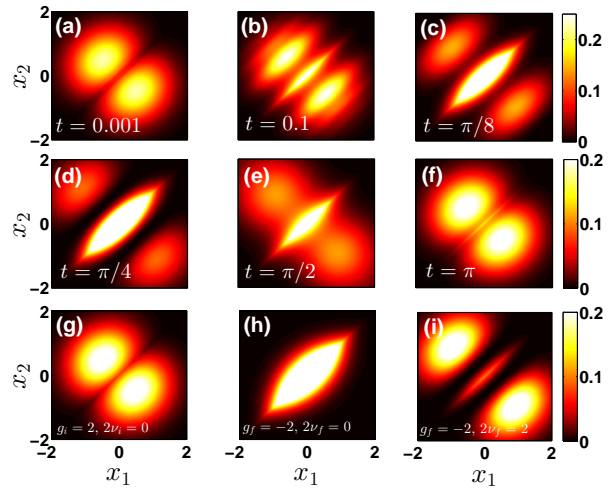}
     \caption{(a)-(f) Time-evolution of the two-body reduced density matrix, $\rho^{(2)}(x_1,x_2;t)$, at different time-instants 
	(see legends) of the dynamics when considering an interaction quench from $g_i=2$ to $g_f=-2$. 
	$\rho^{(2)}(x_1,x_2;t)$ for the ground state of (g) $g_i=2$ and (h) $g_f=-2$ and (i) the second excited state of $g_f=-2$. 
	In all cases the system consists of two harmonically trapped bosons in one-dimension being initialized in their ground state for $g_i=2$.} 
 	\label{fig:5}
\end{figure}

\subsection{Time-Evolution on the Two-Body Level}

To further understand the nonequilibrium dynamics of the two-atoms we subsequently explore the evolution of the two-body reduced density matrix 
$\rho^{(2)}(x_1,x_2;t)$ [see Eq. (\ref{two_body_evol})], illustrated in Figs. \ref{fig:5} (a)-(f). 
At the very early stages of the dynamics $\rho^{(2)}(x_1,x_2;t)$ [see Fig. \ref{fig:5} (a)] resembles $\rho^{(2)}(x_1,x_2;0)$ of the 
initial (ground) state [see Fig. \ref{fig:5} (g)]. 
Indeed, $\rho^{(2)}(x_1,x_2;t)$ possesses a depleted diagonal and its anti-diagonal is more pronounced having a two-hump structure 
which indicates that two-bosons are more likely to reside one in the left and the other in the right side with respect to the 
center $x_1=x_2=0$ of the harmonic oscillator. 
It is worth stressing at this point that the depleted diagonal (also known as correlation hole) of $\rho^{(2)}(x_1,x_2;0)$ stems from the existence of 
the initial strong repulsive interactions \cite{few,momentum}. 
As time evolves, see Figs. \ref{fig:5} (b)-(e), these two-humps located at the anti-diagonal separate in space and 
their amplitude reduces, while the diagonal acquires a finite value being larger than that of the anti-diagonal. 
This latter behavior of $\rho^{(2)}(x_1,x_2;t)$ suggests that in the aforementioned time interval, it is more probable for 
the two-bosons to be found in the vicinity of the center $x_1=x_2=0$ of the harmonic trap than anywhere else in space. 
Moreover the above-described structure of $\rho^{(2)}(x_1,x_2;t)$ is predominantly caused by the involvement of the ground 
and the second excited states in the two-body state of the postquench system, see also Figs. \ref{fig:5} (h), (i). 
For longer times $\rho^{(2)}(x_1,x_2;t)$ returns close to $\rho^{(2)}(x_1,x_2;0)$ of the prequenched ground state [Fig. \ref{fig:5} (f)] and subsequently 
performs a similar to the above-described motion (not shown here) as time increases.

\begin{figure}[ht]
 	\centering
  	\includegraphics[width=0.46\textwidth]{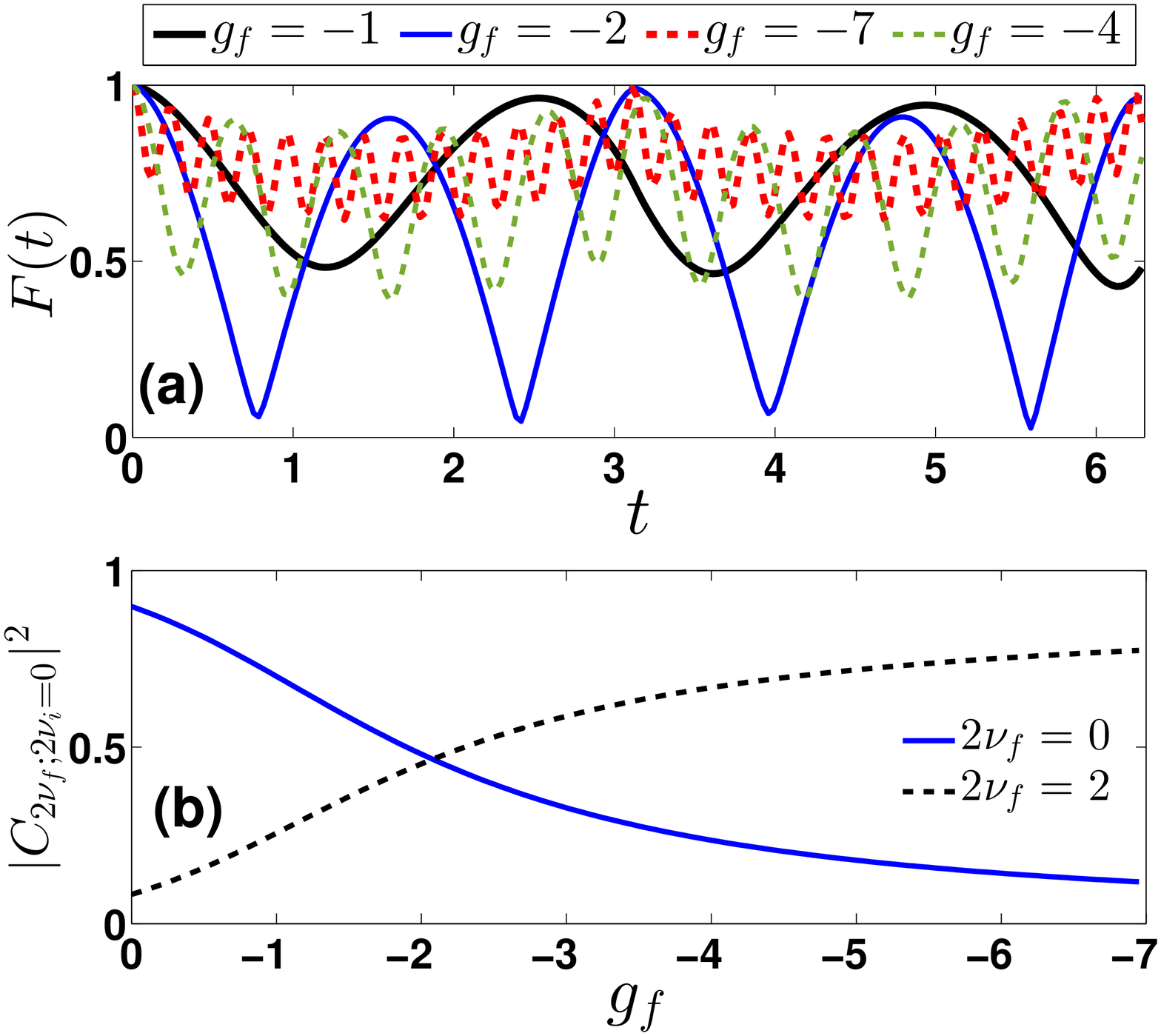}
     \caption{(a) Fidelity evolution after an interaction quench from the ground state of $g_i=2$ to different values of attractive interactions $g_f$ (see legend).
     (b) Square of the overlap coefficients $\abs{C_{2\nu_f;2\nu_i=0}}^2$ between the $2\nu_i=0$ and distinct $2\nu_f$ (see legend) with respect to the postquench interaction strength $g_f$. 
     In both cases, two harmonically trapped bosons in one-dimension are considered.} 
 	\label{fig:F5!}
\end{figure}

\begin{figure}[ht]
 	\centering
  	\includegraphics[width=0.47\textwidth]{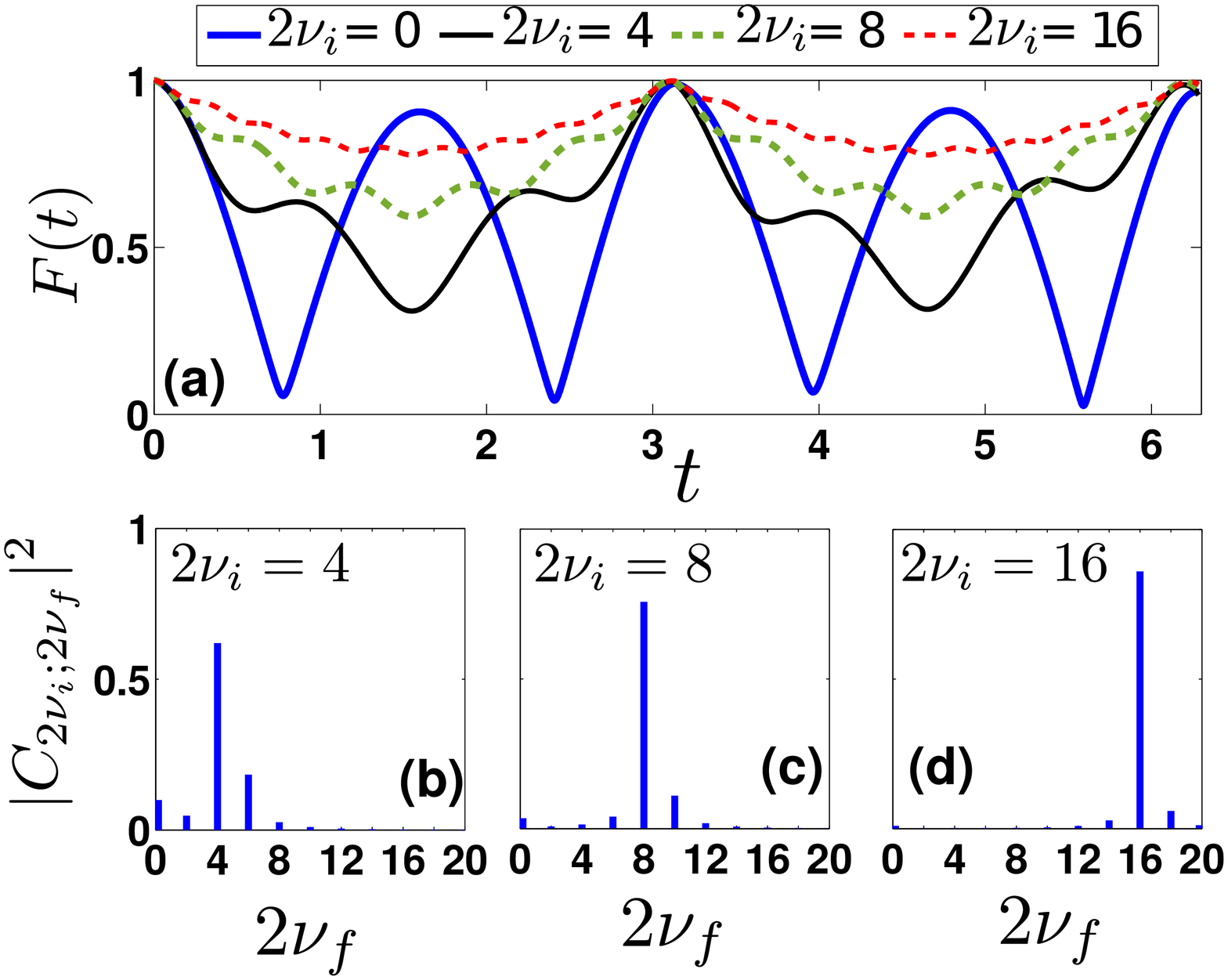}
     \caption{(a) Time-evolution of the fidelity following an interaction quench from a higher-lying excited state (see legend) of the harmonic oscillator for $g_i=2$ to $g_f=-2$. 
      (b), (c), (d) The corresponding $\abs{{C}_{2\nu_f;2\nu_i}}^2$ between an initial even eigenstate $2\nu_i$ (see legend) at $g_i=2$ and different eigenstates of the final, at $g_f=-2$, system. 
      The system consists of two bosons trapped in a one-dimensional hamonic oscillator.} 
 	\label{fig:F5!_1}
\end{figure}

\subsection{Dependence on the Initial and Final States}

Having discussed the characteristics of the nonequilibrium dynamics starting from the ground state of $g_i=2$ and quenching to $g_f=-2$, we subsequently examine the dependence of the 
system's dynamical response on the final interaction strength $g_f$. 
Since we are interested in the behavior of the dynamical response, we invoke as an appropriate measure the fidelity evolution. 
Figure \ref{fig:F5!} (a) presents $F(t)$ for an interaction quench from $g_i=2$ to different negative values of the postquench interaction strength $g_f$. 
It becomes evident that the system's dynamical response can be (roughly) divided into two different interaction regimes, namely one for $g_f>-2$ and the other one for $g_f<-2$. 
Indeed, for $g_f\in (2,-2]$ the oscillation period of $F(t)$ becomes smaller acquiring also a larger amplitude for increasing $g_f$, 
e.g. compare $F(t)$ for $g_f=-1$ and $g_f=-2$. 
On the other hand, when $g_f\in(-2,-\infty)$ $F(t)$ oscillates with both a smaller period and amplitude as $g_f$ takes larger negative values, e.g. see $F(t)$ for 
$g_f=-2.5$ and $g_f=-7$ respectively. 
Let us now interpret the characteristics as well as the origin of the above-mentioned different behavior of $F(t)$ in these interaction intervals. 
It is apparent from $F(t)$ that for an increasing $g_f$ within the interaction interval $(2,-2]$ [$(-2,-\infty)$] the system 
deviates stronger [weaker] from its initial state since the oscillation amplitude of $F(t)$ becomes larger [smaller]. 
This alternating behavior of the oscillation amplitude of $F(t)$ can be understood by inspecting the contribution of the ground and the 
second excited states after the quench, given 
by the overlap coefficients $\abs{C_{2\nu_f;2\nu_i=0}}^2$, for a larger $g_f$. 
Recall that these states are indeed predominantly contributing for the quench under consideration. 
Of course, their magnitudes determine the oscillation amplitude of $F(t)$ [see also Eqs. (\ref{eq:fidelity}) and (\ref{eq:fidelity_spectrum})] which becomes 
maximal when they are comparable. 
Indeed inspecting $\abs{C_{2\nu_f;2\nu_i=0}}^2$ with respect to $g_f$ [Fig. \ref{fig:F5!} (b)] reveals that they possess an equal population at $g_f\approx-2$ while 
for $g_f>-2$ [$g_f<-2$] the postquench ground state has the larger [smaller] occupation. 
To understand the decreasing period of $F(t)$ as $g_f$ increases we determine the energy difference, $\Delta E$, between the ground and the second excited state 
of the postquench system for varying $g_f$, since this $\Delta E$ is mainly responsible for the oscillation period of $F(t)$ [see Eq. (\ref{eq:fidelity_spectrum})]. 
Employing the complete energy spectrum of the two-boson problem presented in Fig. \ref{fig:E} we indeed observe that due to the divergence of the ground state energy branch 
with increasing magnitude of attractive interactions $\Delta E$ increases [see the double arrows in in Fig. \ref{fig:E}]. 
In turn, this increasing tendency of $\Delta E$ results in the corresponding decrease of the oscillation period of $F(t)$. 
\begin{figure}[ht]
 	\centering
  	\includegraphics[width=0.47\textwidth]{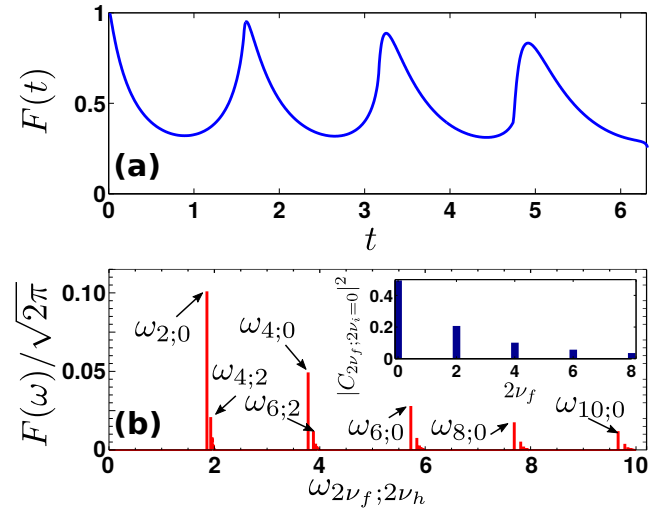}
     \caption{(a) Fidelity evolution after an interaction quench of two harmonically trapped bosons from $g_i=-2$ to $g_{f}=2$ and (b) the corresponding fidelity spectrum. 
     The inset presents the square of the overlap coefficients, $\abs{{C}_{2\nu_f;2\nu_i=0}}^2$, between the initial state ($2\nu_i=0$) of the system and the first 
     five ($\nu'_f\equiv2\nu_f=0,2\dots,8$) even eigenstates of the postquench, at $g_f=2$, system. 
     The two harmonically trapped bosons are initialized in their ground state with $g_i=-2$.} 
 	\label{fig:F6}
\end{figure} 

Next we unravel the dependence of the system's dynamical response on the initial eigenstate of the system for fixed pre- and postquench interaction strengths, namely 
$g_i=2$ and $g_f=-2$. 
The corresponding $F(t)$ is illustrated in Fig. \ref{fig:F5!_1} (a) starting from different excited states. 
As it can be seen, initializing the system in a higher-lying excited state results in a decreasing oscillation amplitude of $F(t)$ but a larger amount of 
involved frequencies, see in particular the increasing number of amplitude oscillations in $F(t)$. 
This behavior can be easily understood by employing the definition of the fidelity [Eq. (\ref{eq:fidelity})] and explicitly determining the 
overlap coefficients, $\abs{C_{2\nu_i;2\nu_f}}^2$ between the initial and the final states 
of the system, see Figs. \ref{fig:F5!_1} (b), (c) and (d). 
Indeed, initializing the system in a higher-lying excited state results in the dominant population of one postquench state whereas also a 
multitude of other states exhibit a very small contribution. 
The participation of all these weakly populated states on the one hand gives rise to a variety of frequencies but also causes a weak 
oscillation amplitude of $F(t)$ since these states possess a very small overlap with the initial state. 
Recall that $F(t)\propto \sqrt{\sum_{2\nu_h<2\nu_f}\cos(\omega_{2\nu_h;2\nu_f})\abs{C_{2\nu_h;2\nu_i}}^2\abs{C_{2\nu_f;2\nu_i}}^2}$, with 
$2\nu_h$, $2\nu_f$ denoting different postquench even eigenstates. 
This is indeed in sharp contrast to the situation in which the system is prepared in its ground state. 
As it has been argued above, in this case the ground and the second excited states mainly contribute to the dynamics [see the inset of Fig. \ref{fig:F5} (b)] 
since their energy gap is smaller when compared to the energy gaps of the postquench ground state with the other states. 
Then $F(t)$ involves a predominant frequency, being the energy difference of these two states, and exhibits a large oscillation amplitude.

\begin{figure}[ht]
 	\centering
  	\includegraphics[width=0.46\textwidth]{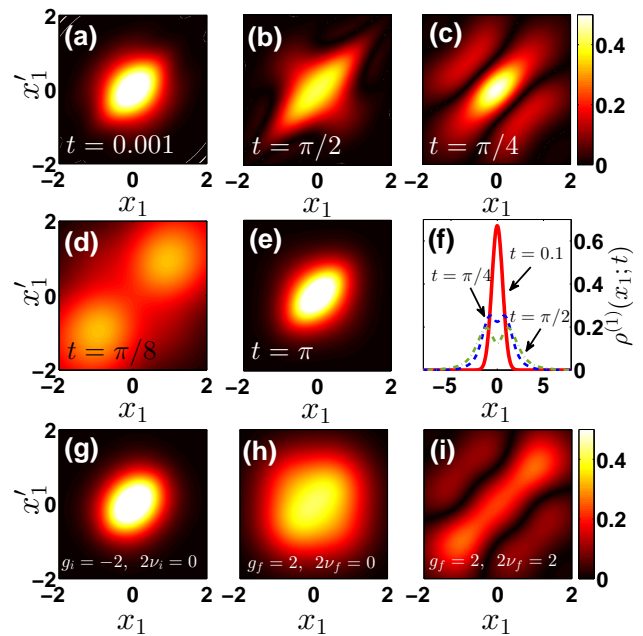}
     \caption{(a)-(e) Snapshots (see legends) of the one-body reduced density matrix, $\rho^{(1)}(x_1,x_1';t)$, following an interaction quench of two initially attractive 
	bosons from $g_i=-2$ to $g_f=2$. 
	(f) Profiles of the one-body density, $\rho^{(1)}(x_1,x_1'=x_1;t)$, for distinct time-instants (see legend) in the course of the evolution. 
	$\rho^{(1)}(x_1,x_1';t)$ for the ground state of (h) $g_i=-2$ and (i) $g_f=2$ and (j) the second excited state of $g_f=2$. 
	In all cases two-bosons are trapped in an one-dimensional harmonic trap and they are prepared in their corresponding ground state with $g_i=-2$.} 
 	\label{fig:SDM6}
\end{figure} 

\begin{figure}[ht]
 	\centering
  	\includegraphics[width=0.46\textwidth]{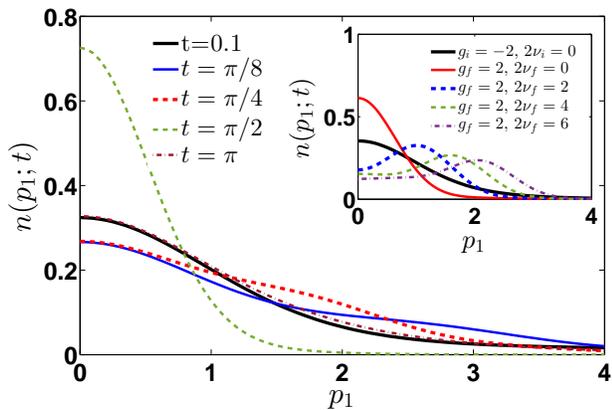}
     \caption{Snapshots of the momentum distribution $n(p_1;t)$ at different time-instants (see legend) of the evolution following an interaction quench from 
	${g}_{i}=-2$ to ${g}_{f}=2$. 
	Only positive momenta ${p}_{1}$ are depicted. 
	The inset illustrates $n(p_1,0)$ for specific system's eigenstates referring to interactions before or after the quench (see legend). 
	The system of two-bosons is initialized in the ground state of an one-dimensional harmonic trap with attractive interactions $g_i=-2$.} 
 	\label{fig:P6}
\end{figure}

\begin{figure}[ht]
 	\centering
  	\includegraphics[width=0.46\textwidth]{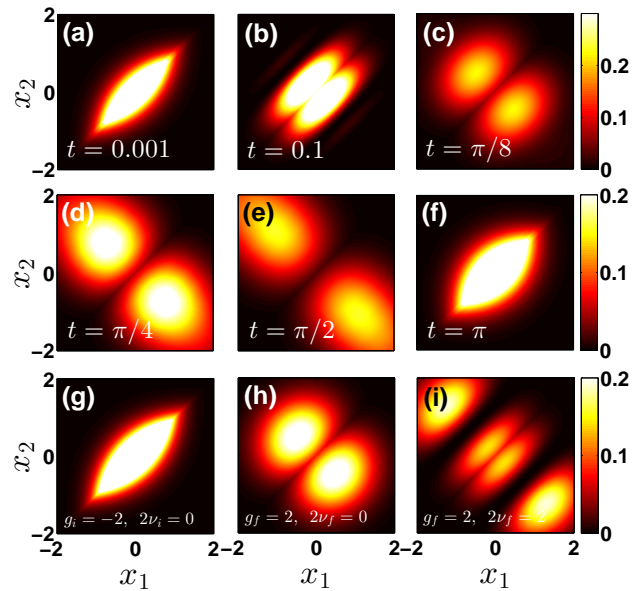}
     \caption{(a)-(f) Dynamics of the two-body reduced density matrix, $\rho^{(2)}(x_1,x_2;t)$, at specific time-instants (see legends) of the evolution when 
	performing an interaction quench from $g_i=-2$ to $g_f=2$. 
	$\rho^{(2)}(x_1,x_2;t)$ for the ground state of (g) $g_i=-2$ and (h) $g_f=2$ and (i) the second excited state of $g_f=2$. 
	In all cases the two harmonically trapped bosons in one-dimension are initialized in their ground state of $g_i=-2$.} 
 	\label{fig:6}
\end{figure}

\section{Quench Dynamics from Attractive to Repulsive Interactions}\label{dynamics_negative}

As a next step we consider the reverse quench scenario, namely quenching the interaction strength from attractive to repulsive values. 
In particular, the system is prepared in its attractively interacting ground state at $g_i=-2$ and we perform an interaction quench 
at $t=0$ to $g_f=2$. 

\subsection{Dynamical Response}

To gain an overview of the system's dynamical response we first employ $F(t)$, shown in Fig. \ref{fig:F6} (a). 
$F(t)$ differs strongly from unity during the evolution, thus showcasing that the system deviates significantly from its initial state. 
Most importantly, $F(t)$ performs oscillations in time with a decaying amplitude. 
The latter behavior suggests that a multitude of frequencies, and therefore different states, contribute to the dynamics. 
Indeed, invoking $F(\omega)$ [Fig. \ref{fig:F6} (b)] we are able to identify the distinct participating frequencies. 
Since these frequencies correspond to transitions between individual states after the quench we also determine the overlap coefficients $\abs{C_{2\nu_f;2\nu_i=0}}^2$, 
between the initial state ($2\nu_i=0$) and the different postquench even eigenstates ($2\nu_f$) of the system. 
$\abs{C_{2\nu_f;2\nu_i=0}}^2$ are shown in the inset of Fig. \ref{fig:F6} (b) for the first five ($2\nu_f=0,2,\dots,8$) even low-lying excited states of the postquench system.  
Evidently, the first three ($2\nu_f=0,2,4$) even lowest-lying states of the system after the quench possess the dominant contribution while the fourth ($2\nu_f=6$) 
and the fifth ($2\nu_f=8$) have a much smaller contribution. 
This knowledge combined with the values of the energy differences between these states (see Table \ref{table}) enables us to assign to each 
peak in $F(\omega)$ the underlying transition between eigenstates. 
Indeed, we can conclude that the dominant peaks appearing in $F(\omega)$ correspond to 
$\omega_{2;0}\approx1.85$, $\omega_{4;0}\approx3.78$, $\omega_{6;0}\approx5.73$, $\omega_{8;0}\approx7.7$ and $\omega_{10;0}\approx9.68$ respectively. 
Note that also other transitions such as $\omega_{4;2}$ and $\omega_{6;2}$ take place exhibiting, however, a smaller contribution [see their amplitude in Fig. \ref{fig:F6} (b)].   

\subsection{One-Body Reduced Density matrix and Momentum Distribution}

To showcase the dynamical spatial redistribution of the atoms on the single-particle level we study the evolution of the one-body reduced density matrix 
$\rho^{(1)}(x_1,x_1';t)$, see Figs. \ref{fig:SDM6} (a)-(f). 
At the initial instants of the evolution, e.g. see Figs. \ref{fig:SDM6} (a) and (b), $\rho^{(1)}(x_1,x_1';t)$ is mainly concentrated in its diagonal exhibiting 
a pronounced peak at $x_1=x_1'=0$. 
This structure, caused by the initial attractive interactions \cite{few_attractive}, highly resembles the ground state distribution of $\rho^{(1)}(x_1,x_1';0)$ [Fig. \ref{fig:SDM6} (g)] 
and indicates that the bosons are mainly localized around the center, $x_1=x'_1=0$, of the trap. 
At later times $\rho^{(1)}(x_1,x_1';t)$ [Figs. \ref{fig:SDM6} (c)-(e)] starts to significantly deform from the previous configuration by means that it becomes more elongated 
along its diagonal while also its off-diagonal elements acquire small values. 
The observed spatial deformation of $\rho^{(1)}(x_1,x_1';t)$ is a manifestation of the superposition of the ground and the second excited 
postquench states [see Figs. \ref{fig:SDM6} (h), (i)] as well as higher-lying excited states (not shown here for brevity reasons) that 
have already been identified via the fidelity spectrum. 
For later times $\rho^{(1)}(x_1,x_1';t)$ comes back very close to its original shape [Fig. \ref{fig:SDM6} (f)] and then for increasing time again deforms (not shown here) 
since the motion is close to periodic, see also $F(t)$. 
Focusing on the diagonal of $\rho^{(1)}(x_1,x_1';t)$, i.e. the one-body density of the system, it shows an expansion and contraction in the course of the evolution 
being a manifestation of the breathing motion of the cloud. 
Regarding the small population of the off-diagonal elements of $\rho^{(1)}(x_1,x_1';t)$ occurring within the time intervals that the system deviates most from its 
initial state [see for instance Figs. \ref{fig:SDM6} (c)-(e) and Fig. \ref{fig:F6} (a)] they essentially indicate the spatial delocalization of a boson. 
\begin{figure}[ht]
 	\centering
  	\includegraphics[width=0.46\textwidth]{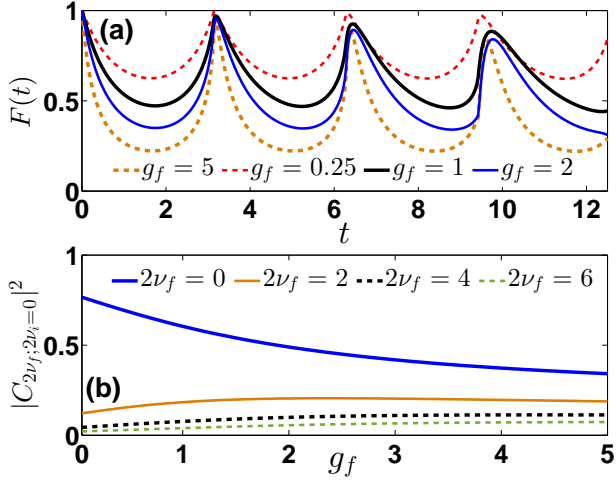}
     \caption{(a) Time-evolution of the fidelity after an interaction quench from the ground state of $g_i=-2$ to different repulsive interaction strengths $g_f$ (see legend). 
	(b) Magnitude of the overlap coefficients $\abs{C_{2\nu_f;2\nu_i=0}}^2$ between $2\nu_i=0$ (initial state) and different $2\nu_f$ final even eigenstates (see legend) 
	for varying postquench interaction strength $g_f$. 
	In both cases, two harmonically trapped bosons in one-dimension are considered.} 
 	\label{fig:F6!}
\end{figure} 

In order to complement our understanding of the dynamics on the one-body level we next employ the momentum distribution, $n(p_1;t)$ \cite{two_atom_ent,Mistakidis_driven}. 
Figure \ref{fig:P6} presents $n(p_1;t)$ in the course of the time-evolution. 
As it can be seen, within the initial stages of the dynamics $n(p_1;t=0.1)$ exhibits a peak at $p_1=0$ and its shape 
almost coincides with that of $n(p_1;0)$ [see the inset of Fig. \ref{fig:P6}]. 
For later times $n(p_1;t)$ becomes broader (e.g. at $t=\pi/4$) and narrower (e.g. at $t=\pi/2$) around $p_1=0$, while its corresponding 
zero-momentum peak takes smaller and higher values respectively. 
This behavior of $n(p_1;t)$ is also indicative of the two-atom breathing motion in momentum space \cite{Jannis,momentum}. 
Another important remark here is that the shape of $n(p_1;t)$ in this time interval is very different from the ground state $n(p_1)$ of $g_f=2$, see the inset of Fig.\ref{fig:P6}. 
Indeed, as we discussed above also higher-lying excited states contribute to the dynamics (especially $2\nu_f=2,4,6$) and 
therefore $n(p_1;t)$ is a superposition of all these states. 
To support our arguments the inset of Fig.\ref{fig:P6} illustrates $n(p_1)$ for the eigenstates $2\nu_f=2,4,6$ of $g_f=2$. 
Notice that the participation of the $2\nu_f>2$ states in the dynamics of $n(p_1;t)$ is much more evident than 
inspecting $\rho^{(1)}(x_1,x_1';t)$. 
For $t=\pi$ the shape of $n(p_1;t)$ tends close to its initial state $n(p_1;0)$ for $g_i=-2$ and then performs the above-described expansion and contraction 
dynamics around $p_1=0$ (not shown).

\subsection{Two-Body Reduced Density Matrix}

The time-evolution of the two-body reduced density matrix, $\rho^{(2)}(x_1,x_2;t)$, provides another spatially resolved measure for the 
nonequilibrium dynamics of the two-bosons, see Figs. \ref{fig:6} (a)-(f). 
For very short evolution times [Fig. \ref{fig:6} (a)] $\rho^{(2)}(x_1,x_2;t)$ shows a bunching tendency across its diagonal. 
More specifically, it has a strong peak in the vicinity of the trap center, i.e. $x_1=x_2=0$, suggesting that it is more likely for two atoms 
to be located in this spatial region. 
This behavior is a consequence of the initial attractive interactions \cite{few_attractive} as it can be deduced by a direct comparison 
with $\rho^{(2)}(x_1,x_2;0)$ [Fig. \ref{fig:6} (g)]. 
As time passes $\rho^{(2)}(x_1,x_2;t)$ exhibits a two-hump structure along its anti-diagonal while the diagonal starts to deplete Fig. \ref{fig:6} (b). 
This two-hump structure is mainly caused by the contribution of the ground and the second excited postquench states [see Figs. \ref{fig:6} (h) and (i) respectively] 
and in part from the higher-lying excited states that are populated during the dynamics. 
Moreover these two-humps spatially separate for later times, see Figs. \ref{fig:6} (b)-(e) and finally merge into the diagonal [Fig. \ref{fig:6} (f)] when the 
system tends close to its initial state as it can be deduced from the fidelity evolution [Fig. \ref{fig:F6} (a)]. 
For later times $\rho^{(2)}(x_1,x_2;t)$ again deforms and undergoes a similar to the above-described dynamics (not shown here). 
\begin{figure}[ht]
 	\centering
  	\includegraphics[width=0.48\textwidth]{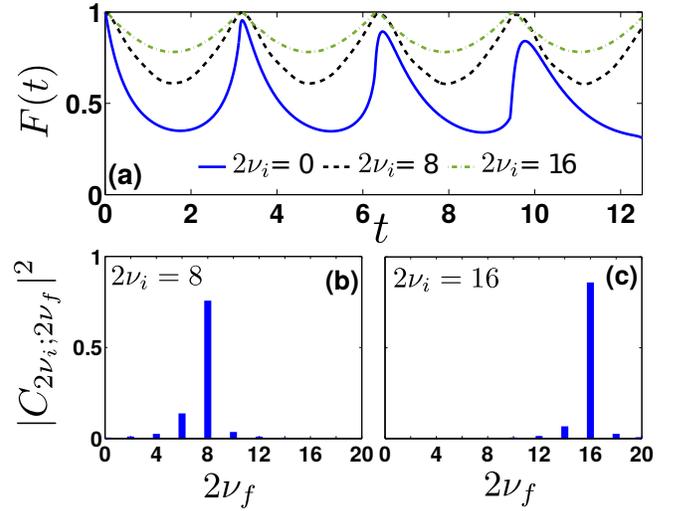}
     \caption{
     (a) Dynamics of the fidelity following an interaction quench from a higher-lying even excited state $2\nu_i$ (see legend) of the harmonic oscillator for $g_i=-2$ to $g_f=2$. 
      (b), (c) The corresponding $\abs{{C}_{2\nu_f;2\nu_i}}^2$ between an initial even eigenstate $2\nu_i$ (see legend) at $g_i=-2$ and distinct even eigenstates of the final system at $g_f=2$.  
      In all cases, two bosons are trapped in a one-dimensional harmonic oscillator. } 
 	\label{fig:F6!_1}
\end{figure} 

\subsection{Further Characteristics of the Dynamical Response}

Next let us examine how the system's dynamical response is affected by the value of the final interaction strength $g_f$. 
To reveal this dependence we determine $F(t)$ for a fixed initial state, being the ground state of the system at $g_i=-2$, and consider interaction 
quenches to different values of a repulsive $g_f$. 
As shown in Fig. \ref{fig:F6!} (a), $F(t)$ differs from unity for every $g_f$ and in particular it exhibits an oscillatory behavior. 
The corresponding oscillation period is almost insensitive to $g_f$. 
This can be understood by the fact that for strong repulsive interactions the energy gaps between the consecutive energy levels are almost constant, 
see also Fig. \ref{fig:E}, and therefore the period of $F(t)$ remains unchanged [see also Eq. (\ref{eq:fidelity})].  
Most importantly, we observe a strong dependence of the oscillation amplitude on $g_f$, suggesting that the system is driven more efficiently 
out-of-equilibrium for stronger repulsive interactions. 
Indeed for increasing $g_f$ this oscillation amplitude becomes larger and for fixed $g_f$ it possesses a decaying tendency in time after 
each oscillation period. 
This latter decaying behavior of the oscillation amplitude in the course of the evolution becomes more prominent for larger $g_f$, e.g. compare $F(t)$ in 
Fig. \ref{fig:F6!} (a) for $g_f=1$ and $g_f=5$. 
Indeed for a stronger repulsive postquench interaction strength $g_f$, more higher-lying excited states acquire a non-negligible amplitude while the corresponding 
ground state becomes less populated, see also Fig. \ref{fig:F6!} (b) where $\abs{C_{2\nu_f;2\nu_i=0}}^2$ is presented for increasing $g_f$. 
The latter fact leads to a larger oscillation amplitude for stronger $g_f$ and a more pronounced dephasing (decaying amplitude) of $F(t)$ since a larger amount 
of states contribute to the dynamics. 

Finally, we investigate the effect of the initial eigenstate of the system on the nonequilibrium dynamics. 
We consider fixed pre- and postquench interaction strengths, i.e. $g_i=-2$ and $g_f=2$ respectively, but initialize the system 
in an energetically different eigenstate. 
Figure \ref{fig:F6!_1} (a) illustrates the time-evolution of $F(t)$ for distinct initial excited eigenstates. 
Overall $F(t)$ shows an oscillatory behavior with a period almost independent of the energetic order of the initial eigenstate. 
However, it is observed that an initially energetically higher eigenstate leads to a smaller oscillation amplitude of $F(t)$ and thus to 
a reduced dynamical response. 
This effect on the oscillation amplitude of $F(t)$ can be understood by inspecting the expansion of the fidelity given by Eq. (\ref{eq:fidelity}). 
Indeed, the deviation of $F(t)$ from unity depends strongly on the values of the overlap coefficients $\abs{C_{2\nu_f;2\nu_i}}^2$ between the initial 
and the final contributing system eigenstates. 
In particular, if a postquench eigenstate possesses a dominant contribution with respect to all others this would result in a smaller oscillation 
amplitude of $F(t)$ when compared to the situation where a multitude of eigenstates are significantly populated. 
The former is exactly the case when the system is prepared in a higher-lying eigenstate [see Figs. \ref{fig:F6!_1} (b), (c)] while the latter is the situation 
where the system is initialized in its ground state [see the inset of Fig. \ref{fig:F6} (b)].

\section{Conclusions}\label{conclusions} 

We have investigated the nonequilibrium quantum dynamics of two harmonically trapped ultracold atoms in one spatial dimension by considering 
an interaction quench from repulsive to attractive interactions and vice versa. 
The interaction potential has been modeled by a contact interaction. 

To set the stage, we provide the analytical expression of the interacting two-body wavefunction 
for an arbitrary stationary eigenstate. 
Moreover, we establish the closed forms of basic observables such as the one- and two-body reduced density matrices, 
the momentum distribution and the fidelity. 
We also briefly discuss the corresponding two-body energy eigenspectrum 
with varying interaction strength ranging from attractive to repulsive values. 
Subsequently, the form of the time-evolving two-body wavefunction is provided. 
In particular, we argue that the corresponding expansion coefficients acquire 
a closed form and therefore the dynamics of the two-body wavefunction can be obtained by 
numerically determining its expansion with respect to the eigenstates of the quenched system. 

Having introduced the theoretical framework for the two-boson system we analyze its dynamics following 
an interaction quench from repulsive to attractive interactions and vice versa. 
To examine the system's dynamical response we utilize the fidelity evolution and its corresponding spectrum. 
This study allows us to identify the predominant participating eigenstates after the quench. 
Next, we unravel the dynamics of the system on both the single- and the two-particle level by inspecting the 
time-evolution of the one- and the two-body reduced density matrices respectively. 
As a consequence of the interaction quench, the system undergoes a breathing motion being visible in the evolution of 
the diagonal of both the one- the two-body density matrices. 

Referring to a quench from repulsive to attractive interactions, it is shown that the anti-diagonal 
of both quantities develops a two-hump structure signaling the involvement of energetically higher-lying states. 
The momentum distribution exhibits a contraction and expansion of its shape in the course of the evolution 
possessing a peak at zero and finite momenta respectively. 
This latter behavior of different populated momenta is another signature of the participation of 
higher-lying eigenstates of the postquench system. 
Remarkably enough, it is shown that the system's dynamical response exhibits a crossover from enhanced to weak response 
as a function of the postquench interaction strength. 
This crossover is found to be related to the crucial participation of the bound state in the postquench 
dynamics for large attractions. 
Finally, we showcase that starting from an energetically lower-lying excited state the system is 
driven more efficiently out-of-equilibrium. 

Turning to the quench from attractive to repulsive interactions 
we show that the energetically higher-lying states that are populated due to the quench are imprinted in the spatial structures 
which develop as time evolves onto the reduced density matrices and in particular along their anti-diagonals. 
Moreover, the breathing motion of the cloud is also evident in the time-evolution of the momentum distribution whose shape 
exhibits strong signatures of the higher-lying populated states. 
Finally, we show that for a fixed initial state of attractive interactions but performing quenches to stronger repulsive interactions results 
in an enhanced dynamical response. 
This behavior holds equally when starting from a lower excited state but considering the same quench amplitude. 

\begin{figure}[ht]
 	\centering
  	\includegraphics[width=0.48\textwidth]{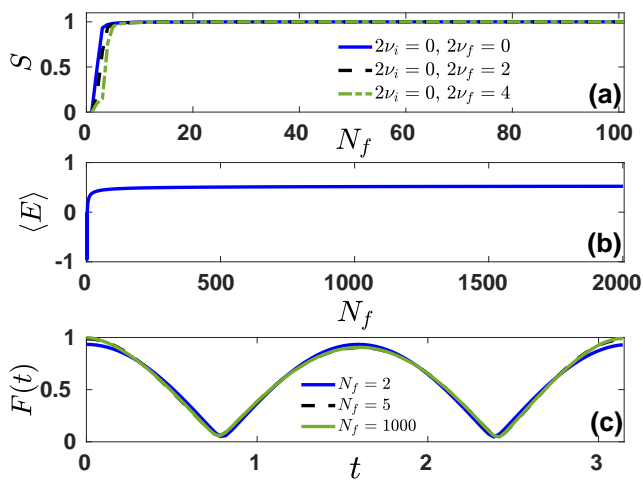}
     \caption{(a) Summation, $S$, over the first $N_f$ states after the quench of the square of the overlap coeficients $\abs{C_{2\nu_f;2\nu_i=0}}^2$ 
     between the pre- ($2\nu_i=0$) and different postquench eigenstates $2\nu_f$ (see legend). 
     (b) Expectation value of the energy over the first $N_f$ final eigenstates after the quench. 
     (c) Fidelity evolution when considering a different number $N_f$ of postquench eigenstates (see legend). 
     In all cases the system consists of two harmonically trapped bosons being prepared in the ground state 
     of $g_i=2$ and we follow an interaction quench to $g_f=-2$.} 
 	\label{fig:C}
\end{figure} 

There are several interesting research directions that can be pursued in a future endeavor. 
A straightforward one would be to consider two-bosons in a two-dimensional harmonic oscillator and examine 
the corresponding interaction quench dynamics from the repulsive to the attractive regime of 
interactions and vice versa. 
Quenches towards the strongly positive or negative scattering lengths (unitarity limit) and the crucial role in the dynamics 
of the existing bound state would be also of particular interest \cite{Corson,Corson1,Corson2}. 
Another intriguing prospect is to investigate the energy spectrum in the dimensional crossover \cite{Corson3} 
from two to one dimensions by considering different trapping frequencies in each spatial direction. 
Subsequently one could utilize these spectra in order to achieve controllable state transfer processes.

\appendix

\section*{Appendix: Numerical Convergence of Observables} \label{convergence}

Let us elaborate on the numerical convergence of our findings. 
Indeed, the time-evolution of all observables is expressed by an expansion over all stationary eigenstates of the postquench system. 
Therefore it is necessary to demonstrate numerical convergence of these observables in terms of the finite basis size that we have used. 
We remark that for our calculations presented in the main text we employ up to the first $10^3$ eigenstates of the postquench system 
depending on the observable of interest (see also below). 
Also, 400 grid points have been used for achieving an adequate spatial resolution of the observables. 

In the following, we showcase the convergence of our results for some representative quantities 
such as the normalization of the considered expansion coefficients, 
the energy of the system and the fidelity evolution after the quench.  
Focusing on the interaction quench from attractive $g_i=-2$ to repulsive interactions $g_f=2$ we show 
below the numerical convergence of the above-mentioned quantities for an increasing number $N_f$ of the system's 
postquench eigenstates. 
Since we consider that the two-body wavefunction, ${\Psi}_{0;2\nu_f}({x}_{1},{x}_{2};t)$ [see also Eq. (\ref{eqn:L})], is normalized to unity then the corresponding 
overlap coefficients should satisfy $\sum_{0 \le \nu_f \le N_f}|{C}_{2\nu_f;2\nu_i}|^2=1$. 
In the latter expression we have introduced a finite upper bound, $N_f$, in the summation indicating the size of our finite truncated basis. 
Figure \ref{fig:C} (a) shows $S\equiv \sum_{0 \le \nu_f \le N_f}|{C}_{2\nu_f;2\nu_i=0}|^2$ of distinct final even eigenstates $2\nu_f$ for increasing $N_f$. 
As it can be readily seen, $S$ converges rapidly for $N_f>30$ independently of the $\nu_f$. 
On the other hand, the expectation value of the energy after the quench reads $\braket{E}=\sum_{0 \le \nu_f \le N_f}{E}_{2\nu_f}|{C}_{2\nu_f;2\nu_i=0}|^2$. 
We observe that $\braket{E}$ [see Fig. \ref{fig:C} (b)] saturates to its final value much slower than $S$ does [compare Figs. (\ref{fig:C}) (a) and (b)] and in particular 
for the specific quench amplitude for $N_f>500$. 
Turning to the fidelity evolution $F(t)=|\sum_{0 \le \nu_f \le N_f} e^{-i{E}_{2\nu_f}t}|{C}_{2\nu_f;2\nu_i=0}|^2|$, see Fig. \ref{fig:C} (c), we can deduce that 
for $N_f>15$ it is insensitive to a further adding of states. 
Finally, we remark that a similar analysis has been performed for all other quench scenarios, e.g. from repulsive to attractive interactions, 
discussed in the main text and found to be absolutely converged (not shown here for brevity).

\section*{Acknowledgements} 
The authors gratefully acknowledge financial support by the Deutsche Forschungsgemeinschaft 
(DFG) in the framework of the SFB 925 ``Light induced dynamics and control of correlated quantum systems''.

{}

\end{document}